\documentclass[useAMS,referee]{biom}

\usepackage{graphicx} 
\usepackage{amsmath,latexsym, graphics,graphicx,showexpl,amsfonts}
\usepackage{graphicx}
\usepackage{subcaption}
\usepackage{bbm} 
\usepackage{setspace}
\let\bibhang\relax
\usepackage{natbib}
\usepackage[titletoc]{appendix}
\usepackage{epstopdf}
\usepackage{pifont}
\usepackage{algpseudocode,algorithm}
\usepackage{multicol}
\usepackage{hyperref}
\usepackage{array}
\usepackage{etoolbox}
\usepackage{setspace}
\usepackage{arydshln}
\usepackage{comment}
\usepackage[table]{xcolor}
\usepackage{multirow}
\usepackage{booktabs}

\title[Frequentist-calibrated Bayesian group sequential design with dynamic borrowing]{Frequentist-calibrated Bayesian group sequential design with dynamic borrowing}

\author{Francesco Mariani$^{1,*}$\email{francesco.mariani2@mail.mcgill.ca},
	Shirin Golchi$^{1}$, Stefania Gubbiotti$^{2}$ and Fulvio De Santis$^{2}$ \\
	$^{1}$Department of Epidemiology and Biostatistics, McGill University, Montr\'eal, Canada\\
	$^{2}$Department of Statistical Sciences, Sapienza University of Rome, Rome, Italy }

\begin{document}
	
	
	\date{{\it Received July} 2026.}
	
	
	
	\pagerange{\pageref{firstpage}--\pageref{lastpage}} 
	\volume{64}
	\pubyear{2008}
	\artmonth{December}
	
	
	\doi{10.1111/j.1541-0420.2005.00454.x}
	
	
	\label{firstpage}
	
	\begin{abstract}
		Bayesian analysis is increasingly used in clinical trials. However, assessment of the design with respect to the frequentist operating characteristics, such as type I error and power, remains a regulatory requirement in many cases. It is well established that, when information is borrowed from external sources to the trial, imposing strict frequentist type I error rate control is equivalent to offsetting the borrowing, which results in no power gains. We provide a framework for frequentist calibration of Bayesian group sequential designs with dynamic borrowing that exploits an explicit correspondence between Bayesian decision criteria based on posterior odds and frequentist uniformly most powerful (UMP) tests. At each interim analysis, two evidential thresholds are made available: the one that exactly retrieves the frequentist UMP decision; the other, that allows the investigator to incorporate historical information when appropriate. We assess the performance of the proposed approach in numerical studies, and apply the framework to the design of a phase III tuberculosis prevention trial, incorporating historical adult and pediatric trial data.
	\end{abstract}
	
	
	\maketitle

	\section{Introduction}
	Clinical trials are increasingly designed and analysed within the Bayesian framework, whose flexibility and capacity for information borrowing are particularly appealing in this context. There remains much room for methodological development, as indicated by the recent draft guidance by the Food and Drug Administration of the United States on the use of Bayesian methodology in clinical trials of drugs and biological products \citep{FDA2026}. Nevertheless, regulatory agencies typically require clinical trial designs to be evaluated with respect to frequentist operating characteristics, namely type I error rate and power. The challenges associated with satisfying frequentist constraints while taking advantage of the Bayesian framework are a recurring theme in many modern clinical trials. These challenges become particularly acute when information is borrowed from historical or external data sources: imposing strict control of the type I error rate when borrowing is employed tends to substantially reduce, if not entirely offset, the inferential gains that borrowing would otherwise provide; see, for instance, \cite{Psioda2019, KoppSchneider2020}. 
	\\
	Bayesian borrowing of historical data has a rich literature spanning several decades, dating back to the seminal work of \cite{Poc76}. A systematic review on the topic is provided by \cite{Wadsworth2018} in the specific context of extrapolating adult data to support pediatric drug development. An important subset of these contributions concerns dynamic borrowing methods, namely approaches that adaptively discount the contribution of historical information based on its concordance with the current data. The practical implications of such methods in the clinical trial setting were systematically explored, for instance, by \cite{Viele14}. More recently, \cite{Lesaffre2024} provided a broad overview of dynamic borrowing methods.
	Despite the large literature on borrowing in the analysis phase of clinical trials, its integration into the design of adaptive experiments has received less attention.
	\\
	Motivated by the design of a phase III Bayesian tuberculosis trial \citep{sstarlettrial},
	we propose a frequentist-calibrated Bayesian group sequential framework with dynamic borrowing that exploits analytic, closed-form relationships between Bayesian decision criteria based on posterior odds and frequentist UMP tests. Rather than attempting to control type I error under borrowing, which would largely defeat its purpose, the framework allows the design to leverage historical information when appropriate, while ensuring that a frequentist counterpart is always available for comparison, when compatibility with past information is questioned, or for transparent communication of the contribution of external data. The key intuition is that Bayesian and frequentist tests can be expressed through a single decision rule, which rejects the null hypothesis when the posterior odds exceed an evidential threshold. Building on the calibration framework of \cite{DSGM2026}, we analytically derive these thresholds which, at each interim stage, allow the investigator to perform either a fully Bayesian or a frequentist UMP test.
	\\
	We note that related analytic approaches to sequential Bayesian decision making based on Bayes factors have also been recently proposed by \citet{Pawel26}; the present work is distinct in that it features designs with dynamic borrowing, while simultaneously introducing an explicit calibration that enables a fully Bayesian test. 
	\\
	The Bayesian-frequentist correspondence has also been exploited in the context of dynamic borrowing by \citet{Calderazzo2026}, who, under a fixed prior, calibrate thresholds on posterior probabilities that adapt to the observed prior-data conflict, allowing type I error rate inflation up to a pre-specified cap, at the cost of moderating the power gains from borrowing. Since posterior odds are a monotone transformation of posterior probabilities, the two approaches operate on equivalent scales; the resulting decision rules, however, differ. In our approach, the threshold for the fully Bayesian test is fixed and it is the prior that adapts. Type I error rate inflation is deliberately accepted as the price for power gain. The inflation in type I error is only limited by the dynamic borrowing mechanism which provides some level of robustness to prior-data conflict. Moreover, we extend the calibration framework to group sequential designs with dynamic borrowing where the amount of historical information incorporated is adaptively reassessed at each interim analysis. The adaptation is made according to a pre-specified rule and based on the concordance between historical and accumulating trial data. At every stage, the threshold that retrieves the frequentist UMP test remains available to quantify or remove the contribution of the prior.
	\\
	In the simulation studies, we highlight an issue specific to dynamic borrowing in group sequential designs: unless the amount of borrowing is adequately bounded relative to the current sample size, historical information plays a greater role in the earlier interim analyses. To protect against potential biases that may be introduced at early stages, the proposed framework is illustrated with a modified version of the meta-analytic predictive (MAP) prior \citep{Neuenschwander10}, in which the amount of borrowing is adjusted according to the similarity and relative size of the historical data compared to the accrued trial data. Note, however, that the proposed Bayesian-frequentist mapping is presented in general terms and can accommodate a range of suitable dynamic borrowing methods, as discussed in Section \ref{sec:db}.
	\\
	The article is structured as follows. Section \ref{sec:motivating} describes the motivating context. Section \ref{sec:methodology} presents the methodology, establishing the correspondence between posterior odds based decision rules and frequentist UMP test statistic and extending it to a group sequential setting with dynamic borrowing. Section \ref{sec:numericalassessments} reports a numerical assessment of the proposed calibration, and Section \ref{subsec:numerical_sstarlet} applies the framework to the SSTARLET trial. Section \ref{sec:discussion} concludes with a discussion.

	\section{The motivating context}
	\label{sec:motivating}
	The proposed methodology in this manuscript is motivated by the design of a phase III Bayesian group sequential trial to establish safety and non-inferiority for tuberculosis preventive treatments (TPT). Tuberculosis (TB) is a major global health concern surpassing COVID-19 as the leading infectious cause of mortality \citep{WHO2025}. While effective TPTs are available, existing regimens have a long treatment period and are poorly tolerated by patients \citep{Dye2013, Uplekar2015}. The SSTARLET (Shorter and Safer Treatment Regimens for Latent Tuberculosis) trial (ClinicalTrials.gov ID: NCT06498414) is a large-scale multi-arm phase II clinical trial that was launched in response to the need for shorter, safer, and more scalable regimens to prevent TB disease. SSTARLET takes an adaptive approach to evaluate multiple outcomes including safety, tolerability and treatment completion for several TPTs allowing for early termination of treatment arms and addition of new arms as new TPTs become available. The decision criteria are defined to allow for a seamless transition to a two arm phase III with the ``shortest and safest" experimental treatment graduating from phase II and the reference treatment is continued. A detailed description of the statistical design of SSTARLET can be found in \cite{Hagar2025}. To facilitate a seamless transition, the planning and design of the Bayesian group sequential phase III trial must occur prior to the completion of the phase II SSTARLET trial. 
	Note that phase II was explicitly designed with identical methods for TB disease ascertainment and diagnosis as phase III to allow direct incorporation of phase II data into the final phase III analysis. 
	However, for illustrative purposes, here we take a dynamic borrowing approach which adjusts the amount of information borrowing based on the accumulating evidence for concordance between the current (phase III) and past data. Beyond the phase II SSTARLET trial itself, additional historical evidence on the reference regimen is available from previously published randomized trials, including adult \citep{study1} and pediatric populations \citep{study2}; simulation studies in Section \ref{subsec:numerical_sstarlet} illustrate how the proposed framework incorporates such heterogeneous sources and remains robust to potential misspecifications.
	

	\section{Methodology}
	\label{sec:methodology}
	Consider the random vector $\boldsymbol{X}$ of $n$ independent observations following a distribution indexed by a vector of parameters $(\theta, \boldsymbol{\xi})$, where $\theta$ is a scalar parameter of interest representing the treatment effect and $\boldsymbol{\xi}$ is a vector of nuisance parameters. 
	Let $\hat \theta(\boldsymbol{X})$ denote an asymptotically normal estimator for $\theta$ such that
	\begin{equation}
		\label{eq:mle}
		\hat{\theta}(\boldsymbol{X}) \stackrel{\cdot}{\sim} \text{N}\left(\theta, \frac{\sigma^2}{n}\right),
	\end{equation}
	where $n$ represents the actual sample size and $\sigma^2 = \sigma^2(\boldsymbol{\xi})$ is the variance that is obtained from the relevant component of the Fisher information matrix with appropriate derivation and depends on the nuisance parameters (see \cite{Spi04}, sec. 2.4). In the remainder of the manuscript, we rely on the above approximate normal distribution.   
	Following \cite{Spi04}, this framework includes, among others, the cases where $\theta$ is log-odds and log-odds ratio for binary outcomes, log hazard ratio for time-to-event outcomes, log rate ratio for count outcomes, and mean difference for continuous outcomes. 
	Let the primary research question be formulated by the hypotheses
	\begin{equation}
		H_0: \theta\leq \theta_0 \quad \mbox{vs} \quad H_1: \theta>\theta_0
		\label{eq:hypotheses}
	\end{equation}
	and let $\hat{\theta}(\boldsymbol{X})$ be the maximum likelihood estimator (MLE) of $\theta$.
	Considering the asymptotically normal test statistic (Wald)
	\begin{equation}
		T(\boldsymbol{X}) = \frac{\sqrt{n} \left(\hat{\theta}(\boldsymbol{X}) - \theta_0 \right)}{\sigma},
		\label{eq:teststatistic}
	\end{equation}
	the uniformly most powerful (UMP) asymptotic test rejects $H_0$ when $T(\boldsymbol{x}) > c_f, \, c_f \in \mathbb{R}$, where $\boldsymbol{x}$ denotes the observed data and $\hat{\theta}(\boldsymbol{x})$ the observed MLE. As $T(\boldsymbol{X})$ is asymptotically normal under $H_0$, for a one-sided frequentist UMP test at level $\alpha$ in a single-stage experiment, the critical value is given as $c_f = z_{1-\alpha}$, where $z_{1-\alpha}$ is the $(1-\alpha)$-quantile of the standard normal distribution.
	
	\subsection{Calibration of the Bayesian decision rule based on posterior odds}
	The standard Bayesian decision rule rejects the null if 
	\begin{equation}
		\label{eq:PO_rule}
		\psi_{10} = \frac{P(\theta>\theta_0\mid \boldsymbol{X})}{P(\theta\leq\theta_0\mid \boldsymbol{X})} > e,
	\end{equation}
	where $\psi_{10}$ is the posterior odds of $H_1$ vs $H_0$ and $e$ is an  evidential threshold; see \cite{KassRaftery} for a discussion. 
	In this section, we describe how to choose $e$ in order to match the Bayesian decision rule with the corresponding optimal frequentist testing procedure. 
	Specifically, we present the one-to-one correspondence between the one-sided frequentist UMP test and the Bayesian decision rule based on the posterior odds, following the ideas in \cite{Shively} and \cite{DSGM2026}. 
	With respect to the normal model introduced in the previous section, 
	assume a normal analysis prior density $\pi(\cdot)$ for $\theta$, with mean $\mu_a$ and variance $\sigma^2/n_a$, where for mathematical convenience we express the prior variance in terms of the observation variance and the effective prior sample size $n_a$.
	Under the hypotheses \eqref{eq:hypotheses}, 
	the posterior odds of $H_1$ vs $H_0$ is
	\begin{equation}
		\label{eq:po}
		\psi_{10} = \psi_\pi(t,n) = \frac{\Phi\left( \frac{\sqrt{n_a} t_a + \sqrt{n}t}{\sqrt{n_a + n}}\right)}{1-\Phi\left(\frac{\sqrt{n_a} t_a + \sqrt{n}t}{\sqrt{n_a + n}}\right)},
	\end{equation}
	where $t = T(\boldsymbol{x})$ and $t_a = \frac{\sqrt{n_a}(\mu_a - \theta_0)}{\sigma}$. 
	Noting that $\psi_\pi(t,n)$ is an increasing function of $t$, equivalence between the rejection rules of the UMP and the Bayesian decision rules is established by the following relationship:
	\begin{equation}
		t > c_f \qquad \iff \qquad \psi_\pi(t,n) > \psi_\pi(c_f,n).    
	\end{equation}
	This mapping allows for specification of the Bayesian decision threshold with the corresponding analysis prior that meets the frequentist operating characteristics implied by the choice of $c_f$.
	Therefore,
	by setting $e = e_f = \psi_\pi(c_f,n)$, the Bayesian decision rule exactly retrieves the frequentist UMP test with critical value $c_f$: the threshold $e_f$ depends on the analysis prior in such a way that its effect cancels out on both sides of the inequality, leaving the rejection rule equivalent to $t > c_f$.
	\\
	Alternatively, to move away from the pure frequentist paradigm, we propose to calibrate $e$ under the non-informative constant analysis prior $\pi_0$ obtained by setting $n_a = 0$, and by replacing $t$ with $c_f$ in \eqref{eq:po}, yielding $e = e_b = \psi_{\pi_0}(c_f,n) = \frac{\Phi(c_f)}{1-\Phi(c_f)}$. 
	The threshold $e_b$ is fixed at the value that makes the Bayesian decision rule equivalent to the frequentist UMP test under a non-informative prior, and it is then kept fixed as the analysis prior is enriched with prior beliefs and/or historical data. When prior information is incorporated, $\psi_\pi(t, n)$ is updated accordingly, while $e_b$ remains anchored to its non-informative value, so that any deviation from the frequentist UMP test is entirely due to the additional information encoded in the prior.
	\\
	In summary, starting from the single unifying rule \eqref{eq:PO_rule}, 
	one can use either the frequentist evidential threshold $e_f$ or
	the non-informative evidential threshold $e_b$:
	the difference between $e_f$ and $e_b$ quantifies the impact of prior information/beliefs
	in the evidential thresholds.
	\\
	As a final remark, note that it is possible to express the Bayesian rejection rule on the scale of $T(\boldsymbol{x})$, by applying the inverse function of $\psi_\pi(\cdot, n)$ to the evidential threshold $e$, that is,
	$$
	\psi_\pi(t,n) > e \qquad \iff \qquad t > \psi_\pi^{-1}(e,n),
	$$
	where 
	\begin{equation}
		\label{eq:inverse}
		\psi_\pi^{-1}(u,n) = \sqrt{\frac{n_a+n}{n}} z_{\frac{u}{1+u}} - \sqrt{\frac{n_a}{n}} t_a.
	\end{equation}

	\subsection{Group sequential design}
	Consider now a group sequential design with $L$ analyses. We denote by $\boldsymbol{x}_{\ell}$ and $n_{\ell}$ the cumulative observed sample and cumulative sample size up to stage $\ell = 1, \dots, L$, and by $\hat{\theta}(\boldsymbol{x}_{\ell})$ the corresponding MLE. 
	From group sequential theory the joint canonical distribution of the test statistics is as follows
	\begin{equation}
		\label{eq:JCD}
		\left( T(\boldsymbol{X}_1), T(\boldsymbol{X}_2)\ldots, T(\boldsymbol{X}_L) \right) ^T\sim MVN(\theta \mathbf{1}_L, \Sigma)
	\end{equation}
	where, from \eqref{eq:teststatistic}, $T(\boldsymbol{X}_\ell) = \frac{\sqrt{n_\ell}(\hat{\theta}(\boldsymbol{X}_\ell) - \theta_0)}{\sigma}$ and where the $(\ell,j)$ generic element of the covariance matrix $\Sigma$ is 
	$\Sigma_{\ell,j}=\sqrt{\frac{n_{\ell}}{n_{j}}}$, for $1\leq \ell\leq j\leq L$ 
	\citep{JennisonTurnbull1997}.
	
	
	\subsubsection{Fixed analysis prior}
	For the moment, consider a fixed analysis prior $\pi(\cdot)$ throughout the trial. The Bayesian decision rule at stage $\ell$ is based on the posterior odds  $\psi_\pi(t_\ell, n_\ell)$, given by \eqref{eq:po} where $t_\ell = T(\boldsymbol{x}_\ell)$. 
	At each stage $\ell$, the rejection rule is 
	$$
	\psi_\pi(t_\ell, n_\ell) > e_{b,\ell},
	$$ 
	where the non-informative evidential thresholds 
	\begin{equation}
		\label{eq:kbl}
		e_{b,\ell} = \frac{\Phi(c_{f,\ell})}{1 - \Phi(c_{f,\ell})},
	\end{equation}
	for $\ell=1, \ldots, L$, are fixed at the beginning of the trial 
	with $c_{f,\ell}$ the frequentist critical value at stage $\ell$. 
	In parallel, for each stage $\ell$, one can also compute the evidential threshold that retrieves the frequentist test, namely 
	\begin{equation}
		e_{f,\ell} = \psi_\pi(c_{f,\ell}, n_\ell),    
		\label{eq:kfl}
	\end{equation} 
	for $\ell=1, \ldots, L$, 
	obtained by evaluating the posterior odds at $c_{f,\ell}$ under $\pi(\theta)$.
	When $\pi(\theta)$ is kept fixed across stages, the thresholds $e_{f,\ell}$ depend on $c_{f,\ell}$ with $\pi(\theta)$ and, therefore, they are fully determined from the beginning of the trial. In contrast, if the analysis prior is updated at each stage using accumulating trial data, then $e_{f,\ell}$ is only determined before the stage $\ell$ data are observed (see Section \ref{sec:db_gsd}).
	\noindent
	Then, the probability of stopping for efficacy at stage $\ell$, having not stopped at any earlier stage, is
	\begin{equation}
		\mathbb{P} \left[\psi_\pi(T_\ell, n_\ell) > e_{b,\ell} \mid \psi_\pi(T_j, n_j) \leq 
		e_{b,j}, \, j = 1, \ldots, \ell-1\right],
	\end{equation}
	where $T_k = T(\boldsymbol{X}_k)$, $k=j,\ell$. These probabilities can be derived as conditionals of the joint canonical distribution given in \eqref{eq:JCD}.
	
	\subsection{Dynamic borrowing within the group sequential design}
	\label{sec:db}
	Suppose now that in addition to data from the current study, information from $H$ historical studies is available. Let $\boldsymbol{Z}_h$ and $\boldsymbol{z}_h$ be the $h$-th random and realized historical samples of size $\nu_h$,  $h = 1, \dots, H$. 
	Assuming that historical information is available through aggregate summaries, we denote by $\hat \theta (\boldsymbol{Z}_h)$ the MLE for $\theta_h$, that is the parameter underlying the $h$-th study, and by $\hat \theta (\boldsymbol{z}_h)$ its observed value. Also, let 
	$$\mathcal{D}_{\tt hist} = \{ (\hat \theta(\boldsymbol{z}_h), \nu_h), \; h=1,\dots, H\}$$ 
	denote the overall collection of historical data summaries, assumed fixed throughout the current trial. For the moment, consider a single-stage design and let 
	$s[\hat \theta(\boldsymbol{z}_h), \hat \theta (\boldsymbol{x})]$ denote the similarity measure between the $h$-th historical data and the current trial data, where $s[\cdot,\cdot]$ is a generic similarity function.
	The closed-form expressions of the calibration framework can be retrieved for a normal analysis prior with mean $\mu_a$ and variance $\sigma^2/n_a$, and considering the normal approximation to the sampling distribution of $T$. 
	We generically express the prior mean $\mu_a$ and the effective sample size $n_a$ as functions of the historical data and, possibly, of the similarity measures based on the current trial data:
	\begin{equation}
		\label{eq:g_n}
		\mu_a = g_\mu(\mathcal{D}_{\tt hist}, \mathcal{S}), \quad n_a = g_n(\mathcal{D}_{\tt hist}, \mathcal{S}),
	\end{equation}
	where $\mathcal{S} = \{s_h, \, h = 1, \dots, H\}$, and where $g_\mu$ and $g_n$ depend on the borrowing method. This is the case, for instance, for power priors, commensurate priors and MAP priors, provided that the parameters governing the amount of borrowing (the power parameter, the commensurability parameter and, for MAP priors, the between-study variability and the weights of the historical studies) are not assigned a prior distribution. 
	Borrowing is static when these parameters are set to constant values, so that $g_\mu$ and $g_n$ depend on $\mathcal{D}_{\tt hist}$ only and the historical information is discounted by a fixed amount, chosen in advance and independently of the current data. Conversely, it is dynamic if they are set as functions of the similarity measures, so that $g_\mu$ and $g_n$ also depend on $\mathcal{S}$. In both cases, the analysis prior remains normal.
	Dynamic borrowing can alternatively be obtained by assigning these parameters a prior distribution; in this case, however, the resulting analysis prior is no longer normal: conjugacy is lost and the closed form in \eqref{eq:po} no longer holds, although the thresholds can still be obtained numerically.
	In what follows, we will focus on similarity-based dynamic priors. In a single-stage design, such priors can only be specified after the current data have been observed, so that the same data would be used both to construct the prior and to perform the test. As this may be criticized as double use of data, we propose an alternative approach that embeds the dynamic prior within a group sequential design where the prior at stage $\ell$ is built on the data collected up to stage $\ell-1$. Besides preserving the closed-form calibration, such a design is also instructive for illustrating the proposed framework where the analysis prior adaptively changes at every stage.

	\subsubsection{Stage-wise adaptation in the group sequential design}
	\label{sec:db_gsd} 
	Let us denote the collection of summary statistics available at the end of stage $\ell$ by
	$$\mathcal{D}_\ell = \{(\hat{\theta}(\boldsymbol{x}_{\ell}),n_\ell)\}, \, \ell = 1, \dots, L$$
	and the collection of similarity measures between the $h$-th historical data and the current trial data collected up to stage $\ell$, by 
	$$
	\mathcal{S}_\ell = \{ s_{h,\ell}, \, h = 1,\dots,H\} , \, \ell = 1, \dots, L
	$$
	where 
	$s_{h, \ell} = s[\hat \theta(\boldsymbol{z}_h), \hat \theta(\boldsymbol{x}_\ell)]$.
	\\
	At each stage $\ell = 2, \dots, L$, the analysis prior mean and effective sample size are given by \eqref{eq:g_n}, with the similarity measure calculated based on the data accumulated up to stage $\ell-1$, denoted by $\mathcal{S}_{\ell-1}$, i.e.,
	\begin{equation}
		\label{eq:g_n_l}
		\mu_{a,\ell} = g_\mu(\mathcal{D}_{\tt hist}, \mathcal{S}_{\ell-1}), \quad n_{a,\ell} = g_n(\mathcal{D}_{\tt hist}, \mathcal{S}_{\ell-1}).
	\end{equation}
	\\
	The design proceeds according to the following steps.
	\begin{enumerate}
		\item At stage $\ell = 1$, no decision/interim assessment is made and data $\mathcal{D}_1$ are collected to start the dynamic borrowing at the subsequent stage $\ell = 2$.
		\item At each stage $\ell \geq 2$, the analysis prior depends directly on historical data and indirectly (through the similarity measure) on data accrued up to stage $\ell-1$, i.e.
		\begin{equation}
			\label{eq:analysis_p}
			\pi(\theta | \mathcal{D}_{\tt hist}, \mathcal{S}_{\ell-1}) = \text{N} \left(\theta \mid \mu_{a,\ell}, \frac{\sigma^2}{n_{a,\ell}}\right), \qquad \ell = 2, \ldots , L,
		\end{equation}
		where $\mu_{a,\ell}$ and $n_{a,\ell}$ are given by \eqref{eq:g_n_l}.
		The posterior distribution is obtained by updating \eqref{eq:analysis_p} with all current data up to stage $\ell$, i.e.  
		$$\pi(\theta \big| \mathcal{D}_{\tt hist}, \mathcal{D}_{\ell}) = \text{N} \left(\theta \Bigg| \frac{n_{a,\ell}\mu_{a,\ell} + n_\ell \hat \theta(\boldsymbol{x}_\ell)}{n_{a,\ell} + n_\ell}, \frac{\sigma^2}{n_{a,\ell} + n_\ell} \right).$$
		
		\item Given the observed data from stage $\ell \geq 2$, according to \eqref{eq:po} we compute 
		\begin{equation}
			\psi_\pi(t_\ell, n_\ell) = \frac{\Phi\left( \frac{\sqrt{n_{a,\ell}} t_{a,\ell} + \sqrt{n_\ell} t_\ell}{\sqrt{n_{a,\ell} + n_\ell}}\right)}{1-\Phi\left(\frac{\sqrt{n_{a,\ell}} t_{a,\ell} + \sqrt{n_\ell} t_\ell}{\sqrt{n_{a,\ell} + n_\ell}}\right)}
			\label{eq:po_GS}
		\end{equation}
		where
		$t_\ell = \frac{\sqrt{n_\ell} (\hat\theta(\boldsymbol{x}_\ell) - \theta_0)}{\sigma}$ and 
		$t_{a,\ell} = \frac{\sqrt{n_{a,\ell}}(\mu_{a,\ell} - \theta_0)}{\sigma}$.
		\item The posterior odds $\psi_\pi(t_\ell, n_\ell)$ is compared either with the non-informative threshold $e_{b,\ell}$ given by \eqref{eq:kbl}
		or with the frequentist threshold $e_{f,\ell}$ given by  \eqref{eq:kfl}.
	\end{enumerate}
	\textbf{Remarks}
	\\
	(i) At each stage $\ell \geq 2$, the analysis prior is specified before the stage $\ell$ data are observed, and the data accrued up to stage $\ell-1$ enter it only through $\mathcal{S}_{\ell-1}$, that is, through the weights given to the historical studies.
	\\
	(ii) The decision at each stage $\ell \geq 2$ is based on the posterior odds computed on all the accrued information from the current trial. Before the trial starts, for each stage $\ell$ the Bayesian evidential thresholds $e_{b,\ell}$ are calculated through \eqref{eq:kbl} from the frequentist critical values $c_{f,\ell}$ (e.g. the O'Brien-Fleming boundaries). In contrast, the frequentist evidential thresholds $e_{f,\ell}$ are computed stage by stage, as they depend on the analysis prior which is updated at the beginning of each stage.
	
	\subsubsection{Example: modified MAP prior}
	\label{sec:MAPexample}
	The functions $g_\mu$ and $g_n$ in \eqref{eq:g_n_l} depend on the specific borrowing method employed. Following, among others, \cite{Spi04, Higgins2009, Neuenschwander10, Schmidli14}, we can assume exchangeability across historical and current data through a hierarchical model, which leads to the MAP prior for $\theta$. To accommodate a stage-wise dynamic borrowing approach, we slightly modify the MAP prior (see Section S1 of the Supplementary Material for details in the single-stage setting). Specifically, for a given stage $\ell$, we set the weight associated with the $h$-th historical study as
	$$
	\omega_{h,\ell} = \frac{s_{h,\ell-1} \nu_h}{ \sigma^2 + s_{h,\ell-1} \nu_h \tau^2_\ell}, \qquad h = 1, \dots, H,
	$$
	where $s_{h,\ell-1} \in [0,1]$ is the similarity between the $h$-th historical study and the current data collected up to stage $\ell-1$, so that $s_{h,\ell-1}\nu_h$ acts as an effective sample size, retaining only a fraction $s_{h,\ell-1}$ of the $h$-th historical data at stage $\ell$, and $\tau^2_\ell$ is the between-study variability at stage $\ell$, whose choice is discussed below. This yields
	\begin{equation}
		\mu_{a,\ell} =g_\mu(\mathcal{D}_{\tt hist}, \mathcal{S}_{\ell-1}) = \frac{\sum_{h = 1}^H \omega_{h,\ell} \hat \theta (\boldsymbol{z}_h)}{\sum_{h=1}^H \omega_{h,\ell}}
		\qquad
		\text{and}
		\qquad
		n_{a,\ell} = g_n(\mathcal{D}_{\tt hist}, \mathcal{S}_{\ell-1}) = \frac{\sigma^2}{ \frac{1}{\sum_{h=1}^H \omega_{h,\ell}} + \tau^2_\ell},
		\label{eq:MAPparam}
	\end{equation}
	where $\mu_{a,\ell}$ is the weighted average of historical data with weights $\omega_{h,\ell}$, $n_{a,\ell}$ is the effective sample size of historical data based on within-study variability $\frac{\sigma^2}{\nu_h}$, between-study variability $\tau^2_\ell$ and similarity measures. The structure of the MAP prior is preserved and the interpretation of $\omega_{h,\ell}$ remains unchanged, with historical sample sizes now replaced by their discounted counterparts $s_{h,\ell-1} \nu_h$. The between-study variability, $\tau^2_\ell$, controls how much historical information is borrowed overall. If $\tau^2_\ell = 0$, then $n_{a,\ell} = \sum_{h=1}^H s_{h,\ell-1} \nu_h$, which can be very large even for a small number of studies; as $\tau^2_\ell \to \infty$, then $\omega_{h,\ell} \to 0$ and $n_{a,\ell} \to 0$, so that historical information is ignored. \cite{Neuenschwander10} show that, regardless of the amount of historical evidence, the prior effective sample size is bounded above by $n_{{\max}} = \frac{\sigma^2}{\tau^2_\ell}$ that is interpreted as the prior maximum sample size. This quantity can be used to elicit $\tau^2_\ell$ in interpretable terms: choosing a desired upper bound on the historical contribution yields $\tau^2_\ell = \frac{\sigma^2}{n_{\max, \ell}}$, where $n_{\max,\ell}$ is the maximum prior effective sample size allowed at stage $\ell$. Since the current trial data should be considered more reliable, we propose to set $n_{\max, \ell}$ equal to the current trial sample size at stage $\ell$, so that the prior effective sample size never exceeds that of the current trial. Thus, as the trial progresses, $\tau^2_\ell$ is updated with the cumulative sample size $n_\ell$, adding another dynamic dimension to the proposed approach. Several choices for the similarity measure are available; for review and comparison, see \cite{Mariani2024}. Among these, we use the Hellinger distance, which has been adopted, for instance, by \citet{Ollier2020} in the context of borrowing for clinical trials. Specifically, we define
	$s_{h,\ell-1} = 1- d[\hat{\theta}(\boldsymbol{x}_{\ell-1}),\hat{\theta}(\boldsymbol{z}_h)]$ where $d[\hat{\theta}(\boldsymbol{x}_{\ell-1}),\hat{\theta}(\boldsymbol{z}_h)]$ is the Hellinger distance between the sampling distributions of the current and the $h$-th historical estimators for $\theta$ and $\theta_h$ ($h = 1, \dots, H$), respectively. We adopt the Hellinger distance mainly for convenience: it takes values in $[0,1]$ and, under the normal approximation \eqref{eq:mle}, has a closed-form expression (see Section S1 of the Supplementary Material). It also depends on the sample sizes of the current and historical studies, so that borrowing is adjusted according to the relative size of the current to historical data. Imbalance in either direction would limit borrowing. We emphasize that other similarity measures and borrowing methods may be accommodated within the frequentist-calibrated group sequential design framework described above.
	\\
	The modified MAP prior is employed in the illustrative example of Section S3 of the Supplementary Material and in the application of Section \ref{subsec:numerical_sstarlet}. In the latter, by comparing the unconstrained choice $\tau^2_\ell = 0$ with $\tau^2_\ell = \sigma^2/n_\ell$, we also show that, when borrowing is not bounded by the current sample size, the rejection rates of the fully Bayesian test can be higher at earlier than at later stages.
	
	\section{Numerical assessments}
	\label{sec:numericalassessments}
	In this section we numerically assess the proposed approach under various fixed analysis prior distributions, representing different scenarios of prior-data agreement. The objective is to provide insight into the interplay between different levels of informativeness of the analysis prior, posterior odds, and calibrated thresholds $e_{f,\ell}$ and $e_{b,\ell}$. This setting is representative of situations where the analyst wishes to incorporate subjective prior beliefs directly into the analysis prior. 
	(For illustration of the dynamic borrowing mechanism using the modified MAP prior of Section \ref{sec:MAPexample}, see the toy example of 
	Section S3 of the Supplementary Material). 
	Consider a group sequential design with $L=3$, conducted at cumulative sample sizes $n_1 =100$, $n_2 = 200$ and $n_3 = 300$. In this case, the same analysis prior is used at each interim look, so that $n_a = n_{a,\ell}$ and $\mu_a = \mu_{a,\ell}$, $\ell = 1,2,3$. We consider four analysis prior settings: flat ($n_a = 0$), moderately informative ($n_a = 10$), informative ($n_a = 30$), and strongly informative ($n_a = 120$), each centered at $\mu_a \in \{-0.3, 0, 0.3\}$ representing varying ``bias". Assume that $\sigma^2 = 4$, $\theta_0 = 0$ and that $c_{f,\ell}$ are the O'Brien-Fleming thresholds. For each analysis prior setting, Table \ref{tab:sensitivity1} reports the thresholds 
	$e_{b,\ell}$ and $e_{f,\ell}$: the former are fixed at the start of the trial; the latter depend on the analysis prior and 
	reproduce the frequentist UMP O'Brien Fleming calibrated test within the Bayesian decision rule.
	
	\begin{table}
		\caption{Values of $e_{b,\ell}$ and $e_{f,\ell}$, $\ell = 1,2,3$, for different analysis priors.}
		\label{tab:sensitivity1}
		\begin{center}
			\resizebox{\textwidth}{!}{%
				\begin{tabular}{ccccccccccc}
					\toprule
					&  & \multicolumn{3}{c}{$\mu_a = -0.3$} & \multicolumn{3}{c}{$\mu_a = 0$} & \multicolumn{3}{c}{$\mu_a = 0.3$} \\
					\cmidrule(lr){3-5} \cmidrule(lr){6-8} \cmidrule(lr){9-11}
					$e$ & $n_a$ & $\ell = 1$ & $\ell = 2$ & $\ell = 3$ & $\ell = 1$ & $\ell = 2$ & $\ell = 3$ & $\ell = 1$ & $\ell = 2$ & $\ell = 3$ \\
					\midrule
					$e_{b,\ell} = e_{f,\ell}$ & 0 & 1454.827 & 60.941 & 21.195 & 1454.827 & 60.941 & 21.195 & 1454.827 & 60.941 & 21.195 \\
					$e_{f,\ell}$ & 10 & 549.109 & 41.494 & 16.600 & 876.402 & 53.514 & 19.950 & 1425.326 & 69.605 & 24.095 \\
					$e_{f,\ell}$ & 30 & 125.061 & 21.420 & 10.679 & 398.555 & 42.574 & 17.849 & 1461.152 & 90.552 & 31.067 \\
					$e_{f,\ell}$ & 120 & 4.794 & 3.060 & 2.451 & 63.587 & 21.085 & 12.155 & 2670.034 & 286.306 & 94.923 \\
					\bottomrule
			\end{tabular}}
		\end{center}
	\end{table}
	
	When a flat analysis prior is used ($n_a = 0$), the two thresholds coincide for each $\ell$. Therefore, the non-informative Bayesian decision rule is equivalent to the frequentist UMP test. When an informative prior is considered, the two thresholds diverge. For fixed $\ell$ and $n_a \neq 0$, $e_{f,\ell}$ increases with $\mu_a$ since more evidence in favour of $H_1$ is needed to match the frequentist decision. For increasing $n_a$ and fixed $\mu_a$, $e_{f,\ell}$ decreases when $\mu_a = -0.3$ and increases when $\mu_a = 0.3$: when the analysis prior strongly favors $H_1$, the posterior odds are generally higher. Therefore, the threshold $e_{f,\ell}$ must be raised accordingly to recover the (more conservative) frequentist decision. Conversely, a prior favoring $H_0$ induces lower posterior odds, requiring a lower threshold. The dependence on the prior of $e_{f,\ell}$ remarks that it is meaningful only as a bridging tool between Bayesian and frequentist decisions, rather than as a standalone Bayesian threshold. In Section S2 of the Supplementary Material, we report the O'Brien-Fleming thresholds $c_{f,\ell}$ used in this analysis, together with the corresponding $c_{b,\ell}$ values, i.e. the frequentist thresholds that retrieve the non-informative Bayesian decision rule under the frequentist decision rule obtained by applying the inverse formula \eqref{eq:inverse}.
	\\
	Consider now three alternative values for the underlying effect, $\theta \in \{-0.3, 0, 0.3\}$. The combinations of values of $\theta$ and $\mu_a$ generate both prior–data agreement and prior–data conflict scenarios, yielding several simulation settings. For each setting, we simulate $M = 10000$ trials. Median values of $\psi_\pi$ are reported in Table \ref{tab:sensitivity2}, whereas Table \ref{tab:sensitivity3} shows values of the empirical power functions, $\tilde \gamma_b(\theta)$ and $\tilde \gamma_f(\theta)$, obtained by comparing the values of $\psi_\pi$ with the appropriate threshold and averaging over simulations. 
	\begin{table}[!h]
		\caption{Median of the posterior odds values, $\tilde{\psi}_\pi$, across $10000$ simulations for several simulation settings at stages $\ell = 1,2,3$.}
		\label{tab:sensitivity2}
		\begin{center}
			\begin{tabular}{ccccccccccc}
				\toprule
				&  & \multicolumn{3}{c}{$\mu_a = -0.3$} & \multicolumn{3}{c}{$\mu_a = 0$} & \multicolumn{3}{c}{$\mu_a = 0.3$} \\
				\cmidrule(lr){3-5} \cmidrule(lr){6-8} \cmidrule(lr){9-11}
				$\theta$ &$n_a$ & $\ell = 1$ & $\ell = 2$ & $\ell = 3$ & $\ell = 1$ & $\ell = 2$ & $\ell = 3$ & $\ell = 1$ & $\ell = 2$ & $\ell = 3$ \\
				\midrule
				&0 & 0.070 & 0.017 & 0.004 & 0.070 & 0.017 & 0.004 & 0.070 & 0.017 & 0.004 \\
				$-0.3$&10 & 0.060 & 0.014 & 0.004 & 0.081 & 0.019 & 0.005 & 0.108 & 0.024 & 0.006 \\
				&30 & 0.045 & 0.011 & 0.003 & 0.102 & 0.024 & 0.006 & 0.214 & 0.047 & 0.012 \\
				&120 & 0.013 & 0.004 & 0.001 & 0.182 & 0.048 & 0.014 & 1.368 & 0.328 & 0.100 \\
				\midrule
				&0 & 0.984 & 0.975 & 0.966 & 0.984 & 0.975 & 0.966 & 0.984 & 0.975 & 0.966 \\
				$0$&10 & 0.784 & 0.827 & 0.844 & 0.986 & 0.976 & 0.967 & 1.238 & 1.151 & 1.107 \\
				&30 & 0.523 & 0.607 & 0.651 & 0.986 & 0.976 & 0.968 & 1.860 & 1.570 & 1.438 \\
				&120 & 0.125 & 0.182 & 0.227 & 0.990 & 0.980 & 0.972 & 7.793 & 5.242 & 4.133 \\
				\midrule
				&0 & 13.688 & 55.729 & 199.529 & 13.688 & 55.729 & 199.529 & 13.688 & 55.729 & 199.529 \\
				$0.3$&10 & 8.935 & 38.187 & 138.864 & 11.874 & 49.112 & 176.780 & 15.971 & 63.696 & 226.478 \\
				&30 & 4.531 & 19.922 & 72.619 & 9.459 & 39.314 & 141.684 & 21.519 & 82.965 & 291.406 \\
				&120 & 0.716 & 2.918 & 9.320 & 5.348 & 19.830 & 66.970 & 74.366 & 263.229 & 890.782 \\
				\bottomrule
			\end{tabular}
		\end{center}
	\end{table}
	
	\begin{table}[!h]
		\caption{Empirical power values of the frequentist test, $\tilde \gamma_f(\theta)$, and of the fully Bayesian test, $\tilde \gamma_b(\theta)$, at stages $\ell = 1,2,3$, across $10000$ simulations, for different analysis priors (prior mean $\mu_a$ and effective sample size $n_a$) and true values of $\theta$. The frequentist test (threshold $e_{f,\ell}$) yields the same power whatever the analysis prior; the fully Bayesian test (threshold $e_{b,\ell}$) coincides with it under the non-informative prior ($n_a = 0$). For $\theta \leq 0$, the values are empirical type I error rates.}
		\label{tab:sensitivity3}
		\begin{center}
			\begin{tabular}{cccccccccccc}
				\toprule
				& & & \multicolumn{3}{c}{$\mu_a = -0.3$} & \multicolumn{3}{c}{$\mu_a = 0$} & \multicolumn{3}{c}{$\mu_a = 0.3$} \\
				\cmidrule(lr){4-6} \cmidrule(lr){7-9} \cmidrule(lr){10-12}
				$\theta$ & Power & $n_a$ & $\ell = 1$ & $\ell = 2$ & $\ell = 3$ & $\ell = 1$ & $\ell = 2$ & $\ell = 3$ & $\ell = 1$ & $\ell = 2$ & $\ell = 3$ \\
				\midrule
				\multirow{5}{*}{$-0.3$} & $\tilde \gamma_f(\theta)$ & any & 0.000 & 0.000 & 0.000 & 0.000 & 0.000 & 0.000 & 0.000 & 0.000 & 0.000 \\
				\cmidrule(lr){2-12}
				& $\tilde \gamma_b(\theta)$ & 0   & 0.000 & 0.000 & 0.000 & 0.000 & 0.000 & 0.000 & 0.000 & 0.000 & 0.000 \\
				&          & 10  & 0.000 & 0.000 & 0.000 & 0.000 & 0.000 & 0.000 & 0.000 & 0.000 & 0.000 \\
				&          & 30  & 0.000 & 0.000 & 0.000 & 0.000 & 0.000 & 0.000 & 0.000 & 0.000 & 0.000 \\
				&          & 120 & 0.000 & 0.000 & 0.000 & 0.000 & 0.000 & 0.000 & 0.000 & 0.000 & 0.000 \\
				\midrule
				\multirow{5}{*}{$0$} & $\tilde \gamma_f(\theta)$ & any & 0.001 & 0.016 & 0.044 & 0.001 & 0.016 & 0.044 & 0.001 & 0.016 & 0.044 \\
				\cmidrule(lr){2-12}
				& $\tilde \gamma_b(\theta)$ & 0   & 0.001 & 0.016 & 0.044 & 0.001 & 0.016 & 0.044 & 0.001 & 0.016 & 0.044 \\
				&          & 10  & 0.000 & 0.010 & 0.034 & 0.000 & 0.014 & 0.042 & 0.001 & 0.019 & 0.048 \\
				&          & 30  & 0.000 & 0.005 & 0.021 & 0.000 & 0.010 & 0.037 & 0.001 & 0.024 & 0.061 \\
				&          & 120 & 0.000 & 0.000 & 0.001 & 0.000 & 0.004 & 0.023 & 0.002 & 0.074 & 0.167 \\
				\midrule
				\multirow{5}{*}{$0.3$} & $\tilde \gamma_f(\theta)$ & any & 0.045 & 0.487 & 0.814 & 0.045 & 0.487 & 0.814 & 0.045 & 0.487 & 0.814 \\
				\cmidrule(lr){2-12}
				& $\tilde \gamma_b(\theta)$ & 0   & 0.045 & 0.487 & 0.814 & 0.045 & 0.487 & 0.814 & 0.045 & 0.487 & 0.814 \\
				&          & 10  & 0.021 & 0.424 & 0.783 & 0.031 & 0.466 & 0.807 & 0.044 & 0.508 & 0.830 \\
				&          & 30  & 0.004 & 0.307 & 0.710 & 0.015 & 0.424 & 0.791 & 0.045 & 0.549 & 0.858 \\
				&          & 120 & 0.000 & 0.032 & 0.326 & 0.001 & 0.274 & 0.721 & 0.074 & 0.748 & 0.947 \\
				\bottomrule
			\end{tabular}
		\end{center}
	\end{table}
	In Table \ref{tab:sensitivity2}, when there is no prior-data conflict ($\theta = \mu_a$), median values of $\psi_\pi$, denoted as $\tilde \psi_\pi$, are well below $1$ for $\theta = -0.3$, close to $1$ for $\theta = 0$, and well above $1$ for $\theta = 0.3$. When prior-data conflict arises, for fixed $\theta = 0.3$ ($\theta = -0.3$), $\tilde \psi_\pi$ decreases (increases) as $\mu_a$ shifts toward $-0.3$ ($0.3$), since the analysis prior inflates the evidence in favor of $H_1$. Nevertheless, $\tilde \psi_\pi$ is greater than $1$ (less than $1$) for $\theta = 0.3$ ($\theta = -0.3$) across all settings, except for the case when $n_a = 120$ for $\ell = 1$. In this case, the analysis prior dominates the likelihood (since $n_1 = 100$), driving the evidence toward $H_0$ even when $\theta = 0.3$ (toward $H_1$ even when $\theta = -0.3$).
	\\
	Table \ref{tab:sensitivity3} shows that, regardless of the analysis prior, the frequentist UMP test achieves the same rejection proportions for a given $\ell$ (rows labelled $\tilde \gamma_f(\theta)$), which depend only on $\theta$ and coincide with those of the fully Bayesian decision rule under the non-informative prior ($n_a = 0$). The fully Bayesian decision rule, on the other hand, adapts to the analysis prior: for fixed $\theta = 0.3$, rejection proportions are lower than their frequentist counterpart when $\mu_a = -0.3$, i.e. under prior-data conflict, and higher when $\mu_a = 0.3$, i.e. under no prior-data conflict, while for $\theta = -0.3$ they are essentially zero under both decision rules. Under the null hypothesis ($\theta = 0$), the frequentist test always controls the type I error rate by design, whereas the fully Bayesian decision rule increases the risk of type I error when the analysis prior strongly favors $H_1$ ($\mu_a = 0.3$, $n_a = 30$).
	\\
	This numerical analysis aligns with the well known fact that the Bayesian approach can either improve or deteriorate performance relative to the frequentist UMP test, depending on the degree of prior-data conflict. This motivates the use of a dynamic borrowing strategy that down-weights historical information based on a compatibility measure, thus allowing one to mitigate the negative effects of potential prior-data conflict, while retaining the benefits of the fully Bayesian approach when prior and data are concordant.
	
	\section{Application to the SSTARLET trial}
	\label{subsec:numerical_sstarlet}
\begin{figure}[!h]
			\centering
			\includegraphics[width=\textwidth]{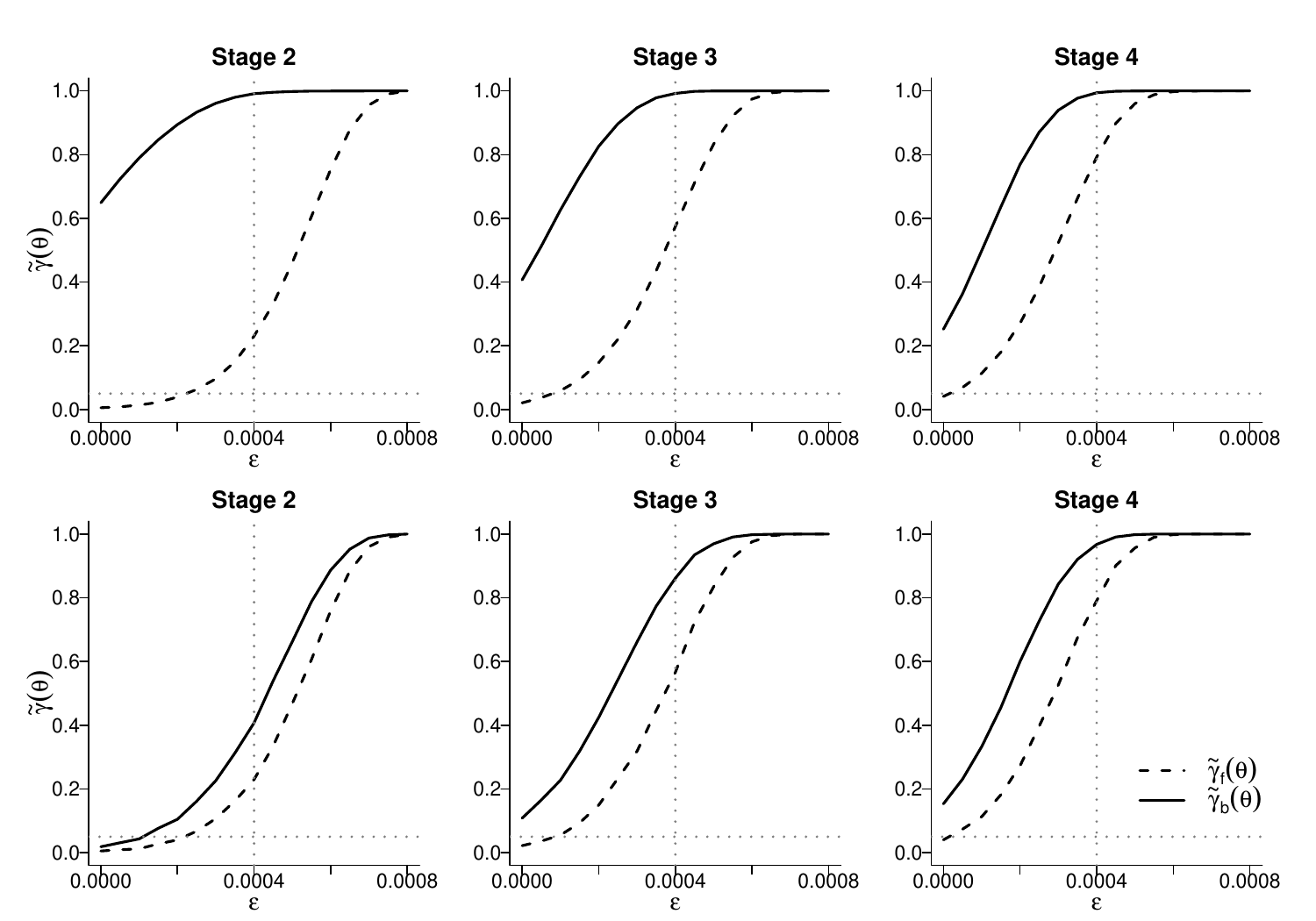}
			\caption{Empirical power curves as a function of $\varepsilon$, for 
				$\tau^2_\ell = 0$ (top panels) and $\tau^2_\ell = \sigma^2/n_\ell$ (bottom panels), 
				at stages $\ell = 2, 3, 4$. Dashed and solid lines correspond to the 
				frequentist and fully Bayesian decision rules, respectively. The vertical 
				gray dotted line is traced at $\varepsilon = 0.0004$, i.e. the hypothesized 
				values $\eta_0 = 0.0010$ and $\eta_1 = 0.6 \times \eta_0$ for which OCs are 
				reported in Table \ref{tab:db_results_SSTARLET}. The horizontal gray dotted 
				line is traced at the nominal type I error rate $0.05$.}
			\label{fig:ERR_curves}
		\end{figure}

	This section presents a simulation study within the context of the SSTARLET trial, assessing the operating characteristics of the proposed adaptive dynamic borrowing approach based on the modified MAP prior of Section \ref{sec:MAPexample}. The hypothetical design comprises three interim analyses ($\ell = 1, 2, 3$) and a final analysis ($\ell = 4$), conducted at cumulative sample sizes $n_1 = 100$, $n_2 = 200$, $n_3 = 300$, and $n_4 = 400$. The primary efficacy outcome is incidence of TB disease. Let $\eta_k, \, k = 0,1$, denote the TB incidence probabilities under the reference and experimental regimens, respectively. We consider the log-odds ratio as the parameter of interest, $\theta = \log\frac{\eta_0/(1-\eta_0)}{\eta_1/(1-\eta_1)}$, which allows using a normal likelihood as an approximate model. 
	As anticipated in Section \ref{sec:motivating}, the design and analysis plan of the phase III trial must be finalized ahead of the completion of the phase II SSTARLET trial; the present simulation study is intended to mimic this hypothetical planning stage.
	Historical information on the incidence of active tuberculosis under the reference treatment (4-month rifampin regimen) of the SSTARLET phase III trial is available from two published randomized trials. In the adult trial reported by \cite{study1}, 4 cases of tuberculosis were observed among 3443 participants assigned to the reference treatment, whereas in the companion pediatric trial reported by \cite{study2}, no cases of active tuberculosis occurred among the 358 children in the same reference treatment. These two sources differ both in the age of the enrolled population and in the precision of the resulting estimates, offering a natural illustration of the proposed dynamic borrowing framework. Note that the 4-month rifampin regimen was the experimental regimen compared against the standard 9-month isoniazid treatment, and now serves as the reference regimen for SSTARLET Phase II and III.
	Since results from the phase II SSTARLET trial are not yet available we consider hypothesized values. 
	Letting $\hat \eta_{kh}$ be the estimated incidence probability under regimen $k$, $k =0,1$, for the $h$-th historical study, 
	we consider $ \hat{\eta}_{01} =  0.0010$ and $\hat{\eta}_{11} = 0.0005$. Summarizing the historical estimates, we have $\hat{\eta}_{02} =0.0012$ from the adult trial and $\hat{\eta}_{03} = 0.0014$ from the pediatric trial. Note that this latter estimate was obtained using a continuity correction, i.e. $\hat{\eta}_{03} = (0+0.5)/(358+1)$, to avoid a degenerate (zero) estimate, since no events were observed in study 3. 
	\\
	Since the parameter of interest is a function of two arm-specific incidence probabilities, and different historical sources are available for each arm, we specify (at each $\ell \geq 2$) separate dynamic MAP priors for $\text{logit}(\eta_0)$ and $\text{logit}(\eta_1)$, which are then combined into a single normal analysis prior for $\theta$. Adopting the strategy of \cite{Spi04} (see Section 2.4.2), we set $\theta_0 = 0$ and $\sigma^2 = 4$. The nominal type I error is set at $0.05$. Full details of the analysis prior construction are provided in Section S4 of the Supplementary Material. 
	\paragraph{Operating characteristics under hypothesized incidence rates}
	The design of the phase III trial is calibrated under the hypothesized incidence rates $\eta_0 = 0.0010$ and $\eta_1 = 0.6 \times \eta_0$, yielding a true log-odds ratio of $\theta = \log\frac{\eta_0/(1-\eta_0)}{\eta_1/(1-\eta_1)} \approx 0.51$. Table \ref{tab:db_results_SSTARLET} reports some Monte Carlo approximations of operating characteristics based on $M=10000$ simulations. Specifically, the table shows the empirical rejection rates for the frequentist and Bayesian procedures, $\tilde \gamma_f(\theta)$ and $\tilde \gamma_b(\theta)$, of the proposed design under the hypothesized effect $\theta=0.51$, for $\tau^2_\ell = 0$ and $\tau^2_\ell = \sigma^2/n_\ell$ at stages $\ell = 2, 3, 4$. It also provides the median posterior odds ($\tilde \psi_\pi$), the median frequentist and Bayesian evidential thresholds ($\tilde e_{f,\ell}$ and $e_{b,\ell}$), the average prior location and effective sample size ($\tilde{\mu}_{a,\ell}$ and $\tilde{n}_{a,\ell}$) and the average arm-specific similarity measure values used at stage $\ell$, i.e. $\tilde s_{h,\ell-1}$ ($\tilde s_h$ for short).
	\begin{table}[!h]
		\caption{Dynamic borrowing results under the hypothesized effect $\theta \approx 0.51$, for $\tau^2_\ell = 0$ and $\tau^2_\ell = \sigma^2/n_\ell$, at stages $\ell = 2, 3, 4$. Symbol $\sim$ denotes quantities obtained as Monte Carlo approximations.}
		\label{tab:db_results_SSTARLET}
		\begin{center}{
					\begin{tabular}{lcccccc}
						\toprule
						& \multicolumn{3}{c}{$\tau^2_\ell = 0$} & \multicolumn{3}{c}{$\tau^2_\ell = \sigma^2/n_\ell$} \\
						\cmidrule(lr){2-4} \cmidrule(lr){5-7}
						& $\ell = 2$ & $\ell = 3$ & $\ell = 4$ & $\ell = 2$ & $\ell = 3$ & $\ell = 4$ \\
						\midrule
						$\tilde \psi_\pi$          & $>10^{14}$ & $>10^{16}$ & $>10^{16}$ & 101.076  & 750.147   & 4591.224  \\
						\midrule
						$\tilde e_{f,\ell}$  & $>10^{15}$ & $>10^{16}$ & $>10^{15}$ & 712.407  & 450.718   & 421.436   \\
						$e_{b,\ell}$                 & 179.390   & 44.670     & 22.419     & 179.390  & 44.670    & 22.419    \\
						\midrule
						$\tilde{\mu}_{a,\ell}$         & 0.801     & 0.794      & 0.788      & 0.837    & 0.824     & 0.814     \\
						$\tilde{n}_{a,\ell}$           & 1391.167  & 1571.010   & 1642.192   & 56.606   & 107.945   & 154.657   \\
						\midrule
						$\tilde{s}_{1}$ (arm 0)        & 0.265     & 0.326      & 0.367      & 0.265    & 0.326     & 0.367     \\
						$\tilde{s}_{2}$ (arm 0)        & 0.217     & 0.250      & 0.267      & 0.217    & 0.250     & 0.267     \\
						$\tilde{s}_{3}$ (arm 0)        & 0.349     & 0.336      & 0.306      & 0.349    & 0.336     & 0.306     \\
						$\tilde{s}_{1}$ (arm 1)        & 0.247     & 0.280      & 0.293      & 0.247    & 0.280     & 0.293     \\
						\midrule
						$\tilde \gamma_f(\theta)$                      & 0.231     & 0.574      & 0.792      & 0.231    & 0.574     & 0.792     \\
						$\tilde \gamma_b(\theta)$                      & 0.991     & 0.992      & 0.994      & 0.408    & 0.862     & 0.967     \\
						\bottomrule
				\end{tabular}}
			\end{center}
		\end{table}
		The two $\tau^2_\ell$ specifications differ substantially in the degree of historical borrowing, as reflected by values of $\tilde{n}_{a,\ell}$. When $\tau^2_\ell = 0$, the prior effective sample size is unconstrained and grows proportionally to the discounted historical sample sizes: since the hypothesized phase II trial (study $1$) and the adult trial (study $2$) both contribute very large historical sample sizes, $\tilde{n}_{a,\ell}$ already exceeds $1391$ at $\ell = 2$ and it keeps increasing at subsequent stages, remaining much larger than the current trial sample sizes. As a result, the borrowing regime is extremely aggressive, the posterior is dominated by historical information rather than by the accruing data, and the Bayesian power $\tilde \gamma_b(\theta)$ is already close to $1$ from the first interim stage, substantially exceeding the frequentist power $\tilde \gamma_f(\theta)$ at all stages. While these results are numerically favorable, they reflect an excessive degree of borrowing that is not adequately restrained by the data. When instead $\tau^2_\ell = \sigma^2/n_\ell$, the prior effective sample size is bounded above by $n_\ell$, yielding much more moderate values of $\tilde{n}_{a,\ell}$ (ranging from about $57$ to $155$ across stages) and correspondingly more measured, though still meaningful, gains of $\tilde \gamma_b(\theta)$ over $\tilde \gamma_f(\theta)$. Among the two specifications these results are considered most satisfactory, as they preserve the benefit of borrowing without allowing historical information to dominate the analysis.
		\\
		The average arm-specific similarity measures $\tilde{s}_h$ are identical across the two $\tau^2_\ell$ specifications at each stage, since $\tau^2_\ell$ does not enter the Hellinger distance. Within the reference arm, the Hellinger distance discriminates meaningfully between historical studies of different compatibility with the hypothesized $\eta_0 = 0.0010$. From the discussion in Section \ref{sec:db}, it follows that a desired property of the Hellinger distance is that it captures similarity jointly in terms of point estimates and sample sizes, rather than through the point estimates alone. The adult trial (study $2$) illustrates this feature clearly: although its observed incidence rate closely matches the hypothesized value, its very large sample size ($3443$) makes it substantially more precise than the current trial data, and this precision mismatch is penalized by the similarity measure, yielding the lowest similarity throughout stages. The pediatric trial (study $3$), with a much smaller sample size ($\nu_3 = 358$) but a point estimate somewhat more distant from the hypothesized value, is initially the most compatible source, but it is progressively penalized as the current sample size grows. The hypothesized phase II trial, with a moderately large sample size and a point estimate that coincides with the hypothesized $\eta_0$, shows the opposite pattern, with $\tilde{s}_1$ increasing from $\approx 0.27$ at $\ell = 2$ to $\approx 0.37$ at $\ell = 4$, eventually becoming the most similar source. Within the experimental arm, where the phase II trial is the only historical source, the similarity $\tilde{s}_1$ follows a comparable increasing pattern (from $\approx 0.25$ to $\approx 0.29$).
		\paragraph{Operating characteristics as a function of the true effect}
		To assess robustness to misspecification of the hypothesized effect, we let $\eta_1 = \eta_0 - \varepsilon$ with $\varepsilon \in [0, 0.0008]$, so that $\varepsilon = 0$ corresponds to the null hypothesis and increasing values of $\varepsilon$ move toward the alternative. We compute the empirical powers $\tilde \gamma_f(\theta)$ and $\tilde \gamma_b(\theta)$ at values of $\theta = \log \frac{\eta_0 /(1-\eta_0)}{\eta_1 / (1-\eta_1)}$ obtained for $\varepsilon \in [0, 0.0008]$: at $\varepsilon = 0$ these reduce to the empirical type I error rate, whereas for $\varepsilon > 0$ they represent the empirical power under the frequentist and fully Bayesian decision rules, respectively. The resulting curves are displayed in Figure \ref{fig:ERR_curves} for both $\tau^2_\ell = 0$ (top panels) and $\tau^2_\ell = \sigma^2/n_\ell$ (bottom panels), at stages $\ell = 2, 3, 4$. The vertical dotted line marks the hypothesized values of Table \ref{tab:db_results_SSTARLET} (i.e., $\eta_0 = 0.0010$, $\eta_1 = 0.6 \times 0.0010$ and $\varepsilon = 0.0004$), and the horizontal dotted line marks the nominal level $0.05$.
		As shown in Figure \ref{fig:ERR_curves}, the type I error rate inflation of the Bayesian decision rule, visible at $\varepsilon = 0$, is unavoidable whenever historical information is borrowed under prior-data conflict, since $\eta_1$ under $H_0$ coincides with $\eta_0$ but the historical sources remain anchored to the hypothesized alternative. The frequentist test, by contrast, preserves the nominal level $0.05$ at $\varepsilon = 0$ at every stage, as expected by construction. When $\tau^2_\ell = 0$, inflation of type I error rate for the Bayesian rule is severe at all stages, mirroring the aggressive, unconstrained borrowing that was discussed earlier: the same lack of restraint that drives $\tilde \gamma_b(\theta)$ far above $\tilde \gamma_f(\theta)$ under the alternative ($\varepsilon = 0.0004$) also drives a substantial inflation of the type I error rate at $\varepsilon = 0$. Also, the rejection rate of the fully Bayesian decision rule is highest at the earliest stage and decreases thereafter. This is because the ratio between the prior effective sample size and the current sample size is largest at the early stages, during which the prior dominates the data more (see Table~\ref{tab:db_results_SSTARLET}, where $\tilde{n}_{a,\ell}$ are displayed for $\varepsilon = 0.0004$). The same mechanism underlies the type I error rate inflation observed at $\varepsilon = 0$, which decreases as $\ell$ increases. When instead $\tau^2_\ell = \sigma^2/n_\ell$, the power gain of the Bayesian test over the frequentist test under the alternative is smaller than for $\tau^2_\ell = 0$, as well as the type I error rate inflation, which remains much closer to the nominal level.
		This trade-off confirms that $\tau^2_\ell = \sigma^2/n_\ell$ offers a more balanced compromise between the potential benefits and the risks of dynamic borrowing which is consistent with the conclusions drawn from Table \ref{tab:db_results_SSTARLET}.
		
		\noindent
		Overall, under the alternative ($\varepsilon > 0$), the fully Bayesian decision rule yields substantial gains in power over the frequentist test at all stages, with the most pronounced differences at earlier stages, when the current data are less informative and the prior contribution is relatively larger. When $\tau^2_\ell = 0$, borrowing is extremely aggressive and the gains are remarkable, with $\tilde \gamma_b(\theta)$ far exceeding $\tilde \gamma_f(\theta)$ across the entire alternative region. However, this comes at a cost: under the null ($\varepsilon = 0$), the type I error rate is very high as a consequence of unconstrained borrowing. When instead $\tau^2_\ell = \sigma^2/n_\ell$, the prior effective sample size is bounded above by $n_\ell$, which moderates the borrowing. The type I error rate remains only slightly above the nominal level, while the power gains over the frequentist test are still remarkable.
		\\
		\paragraph{Robustness to exchangeability violations}
		We consider two additional available historical studies that investigated rifapentine-based regimens of shorter duration than the 4-month rifampin reference, representing a mismatch in the intervention and thus offering an illustration of the robustness of the proposed Bayesian approach when the exchangeability assumption may be violated. In the PREVENT TB trial \citep{study3}, 3 months of rifapentine plus isoniazid were administered, and $7$ cases of confirmed tuberculosis were observed among $3986$ participants assigned to this arm. In the BRIEF TB/A5279 trial \citep{study4}, 1 month of rifapentine plus isoniazid was administered to HIV-infected adults, and $18$ confirmed cases of tuberculosis were observed among $1488$ participants assigned to this arm. Unlike the previous sources, neither trial evaluates the exact reference regimen of the SSTARLET phase III trial, and the latter also enrolls a substantially higher-risk population (HIV-infected adults); both discrepancies are expected to be captured by the Hellinger similarity measure. The corresponding historical estimates are $\hat{\eta}_{04} = 7/3986 \approx 0.0018$ for the PREVENT TB trial and $\hat{\eta}_{05} = 18/1488 \approx 0.0121$ for the BRIEF TB/A5279 trial.
		Table \ref{tab:db_results_SSTARLET_5studies} shows that the similarity measures for these two additional sources are close to zero at every stage, meaning that the use of the Hellinger distance effectively protects the analysis from violations of exchangeability. Consequently, the operating characteristics coincide with those of the more realistic case reported in Table \ref{tab:db_results_SSTARLET}.
		
		\begin{table}[!h]
			\caption{Dynamic borrowing results under the hypothesized effect $\theta = 0.51$, including the two additional historical sources (PREVENT TB and BRIEF TB/A5279), for $\tau^2_\ell = 0$ and $\tau^2_\ell = \sigma^2/n_\ell$, at stages $\ell = 2, 3, 4$. Symbol $\sim$ denotes quantities obtained as Monte Carlo approximations.}
			\label{tab:db_results_SSTARLET_5studies}
			\begin{center}{
						\begin{tabular}{lcccccc}
							\toprule
							& \multicolumn{3}{c}{$\tau^2_\ell = 0$} & \multicolumn{3}{c}{$\tau^2_\ell = \sigma^2/n_\ell$} \\
							\cmidrule(lr){2-4} \cmidrule(lr){5-7}
							& $\ell = 2$ & $\ell = 3$ & $\ell = 4$ & $\ell = 2$ & $\ell = 3$ & $\ell = 4$ \\
							\midrule
							$\tilde \psi_\pi$          & $>10^{18}$ & $>10^{18}$ & $>10^{17}$ & 131.336  & 1102.922  & 6755.365  \\
							\midrule
							$\tilde e_{f,\ell}$  & $>10^{19}$ & $>10^{17}$ & $>10^{16}$ & 978.792  & 669.425   & 567.102   \\
							$e_{b,\ell}$                 & 179.390   & 44.670     & 22.419     & 179.390  & 44.670    & 22.419    \\
							\midrule
							$\tilde{\mu}_{a,\ell}$         & 0.892     & 0.843      & 0.816      & 0.925    & 0.895     & 0.864     \\
							$\tilde{n}_{a,\ell}$           & 1484.390  & 1631.071   & 1680.973   & 58.252   & 110.694   & 157.766   \\
							\midrule
							$\tilde{s}_{1}$ (arm 0)        & 0.265     & 0.326      & 0.367      & 0.265    & 0.326     & 0.367     \\
							$\tilde{s}_{2}$ (arm 0)        & 0.217     & 0.250      & 0.267      & 0.217    & 0.250     & 0.267     \\
							$\tilde{s}_{3}$ (arm 0)        & 0.382     & 0.401      & 0.392      & 0.382    & 0.401     & 0.392     \\
							$\tilde{s}_{4}$ (arm 0)        & 0.110     & 0.066      & 0.038      & 0.110    & 0.066     & 0.038     \\
							$\tilde{s}_{5}$ (arm 0)        & 0.000     & 0.000      & 0.000      & 0.000    & 0.000     & 0.000     \\
							$\tilde{s}_{1}$ (arm 1)        & 0.247     & 0.280      & 0.293      & 0.247    & 0.280     & 0.293     \\
							\midrule
							$\tilde \gamma_f(\theta)$                      & 0.231     & 0.574      & 0.792      & 0.231    & 0.574     & 0.792     \\
							$\tilde \gamma_b(\theta)$                      & 0.993     & 0.993      & 0.994      & 0.450    & 0.879     & 0.969     \\
							\bottomrule
					\end{tabular}}
				\end{center}
			\end{table}

			\section{Discussion}
			\label{sec:discussion}
			We propose a Bayesian group sequential design with dynamic borrowing that links Bayesian decision rules with frequentist UMP testing through calibrated evidential thresholds: a historically informed threshold, $e_{f,\ell}$, that exactly recovers the frequentist decision, and a non-informative threshold, $e_{b, \ell}$, allowing one to move away from the frequentist paradigm and exploit the benefits of borrowing in a Bayesian fashion.
			\\
			While we present the calibration approach in the posterior odds scale, by \eqref{eq:inverse}, both evidential thresholds can be expressed on the scale of the test statistic and, more generally, on that of any monotone transformation of the posterior odds, such as posterior probabilities or Bayes factors.
			\\
			The proposed group sequential borrowing strategy is generic, and specific modelling choices, such as the use of a MAP prior, the specification of the between-trial heterogeneity, and the Hellinger distance as a similarity measure, may be replaced by suitable alternatives.
			\\
			Numerical analyses confirm the well-known finding that the fully Bayesian approach can either improve or deteriorate performance relative to the frequentist UMP test, depending on the degree of prior-data conflict. In the presence of historical data, this motivates the use of a dynamic borrowing strategy, which provides some level of protection against prior-data conflict while retaining the benefits of the fully Bayesian approach when prior and data are in agreement. In the SSTARLET trial application, we further illustrate the importance of bounding the amount of borrowing: employing the modified MAP prior described in Section~\ref{sec:MAPexample}, we show that setting $\tau^2_\ell = \sigma^2/n_\ell$ ensures that the prior effective sample size never exceeds the current trial sample size, so that the prior never dominates the current trial evidence. This offers a more satisfactory compromise than an unconstrained choice, as it moderates the inflation of type I error rate while preserving power gains. Moreover, in the specific context of group sequential designs, constrained borrowing prevents the counterintuitive behaviour, observed under unconstrained borrowing, whereby the rejection rate of the fully Bayesian test is higher at earlier than at later stages. This issue is crucial in group sequential designs: since current information varies across stages while historical information remains fixed, an unconstrained prior can overpower the data at the earliest interim analyses.
			Connecting the bound on the prior effective sample size to $n_\ell$ preserves the balance between historical and current information at every analysis. The application also shows that similarity-based discounting effectively down-weights historical sources that violate the exchangeability assumption, as illustrated when two additional, less comparable regimens were introduced among the historical sources.
			\\
			Our approach also overcomes a related pitfall noted by \cite{Harun2020}: ``no-borrowing'' designs are often implicitly calibrated on assumed historical effects, whereas our $e_{b,\ell}$ is by construction free of any historical-data dependence.

			\section*{Funding}
			SG acknowledges support by the Natural Sciences and Engineering Research Council of Canada (NSERC) and the Fonds de recherche du Québec – Nature et technologies (FRQNT).
			
			\section*{Acknowledgments}
			The authors thank Dr. Dick Menzies for valuable input regarding the hypothetical design exercise in Sections~\ref{sec:motivating} and~\ref{subsec:numerical_sstarlet} motivated by the SSTARLET trial.
			\\
			We used Claude for the following purposes: assistance with ShinyApp development and writing assistance. The authors take full responsibility for the final content of the manuscript and have reviewed all AI-generated material.
			
			\section*{Supplementary Materials}
			The manuscript is accompanied by complementary results and discussions available in the Supplementary Material. To ensure reproducibility of results and implementation of the proposed method, we make the R code available at \\\url{https://github.com/framar1997/FcalibratedBGSD.git}.
			Finally, we supply a ShinyApp available at 
			\url{https://6kp5ow-francesco-mariani.shinyapps.io/FcalibratedBGSDApp/}.
			\\
			Details and instructions on the ShinyApp are provided in Section S5 of the Supplementary Material.
			
			\section*{Conflict of interest}
			None declared.

			\bibliographystyle{oup-abbrvnat}
			\bibliography{Biblio5.bib}
			
			\label{lastpage}
			
		\end{document}